\documentclass[prl,aps,twocolumn,amssymb]{revtex4} 
\usepackage{epsfig,amsmath}
\usepackage{psfrag}

\begin{document}

\title{Synchronization of two interacting populations of oscillators}

\author{Ernest Montbri\'o, J{\"u}rgen Kurths and Bernd Blasius}
\affiliation{Institut f\"ur Physik, Universit\"at Potsdam,
Postfach 601553,D-14415 Potsdam, Germany}
\date{\today}

\begin{abstract}

We analyze synchronization between two interacting populations of different phase oscillators. For the important case of asymmetric coupling functions, we find a much richer dynamical behavior compared to that of symmetrically coupled populations of identical oscillators \cite{OK91}. It includes three types of bistabilities, higher order entrainment and the existence of states with unusual stability properties. All possible routes to synchronization of the populations are presented and some stability boundaries are obtained analytically. The impact of these findings for neuroscience is discussed.

\end{abstract}
\maketitle 

In its original sense synchronization describes the mutual adjustment of frequencies between two interacting oscillators \cite{PRK01}. This route to synchronization differs from that taking place in large communities of oscillators \cite{Win80}. Motivated by this fact, Kuramoto extended the notion of synchronization to a statistical theory of oscillator ensembles \cite{Kur84}. The natural diversity among the components was considered through either unimodal \cite{Kur84,SM91,Sak88} or bimodal \cite{BNS92,Kur84} frequency distributions. 


In a pioneering work, Okuda and Kuramoto \cite{OK91} analyzed two symmetrically coupled populations of identical phase oscillators under the influence of noise. However, the important problem of synchronization between ensembles of oscillators remains almost unexplored, although communities of natural oscillators are usually composed of interacting subpopulations \cite{Win80}. For instance, it has been shown experimentally that synchronization arises between different neighboring visual cortex columns and also between different cortical areas, where synchronization processes are of crucial importance \cite{GKE+89}. Thus, it is a challenge to understand quantitatively the routes to synchronization among macroscopic ensembles of oscillators. 

In this letter we study two populations of phase oscillators interacting asymmetrically, as it is likely to occur due to a number of reasons (asymmetric couplings, different population sizes, time delays...). In addition, we consider the oscillators within each population to be nonidentical. The system under study is then
\begin{equation}
\begin{split}
\dot{\theta}_i^{(1,2)} = &  \omega_i^{(1,2)} - \frac{K_p}{N} \sum_{j=1}^{N} \sin (\theta_i^{(1,2)}-\theta_j^{(1,2)}+\alpha) \\& - \frac{K}{N} \sum_{j=1}^{N} \sin (\theta_i^{(1,2)}-\theta_j^{(2,1)}+\alpha),
\end{split}
\label{model0}
\end{equation}
where $i=1,\dots ,N \gg 1$. Here, $\theta_i^{(1,2)}$ describes the phase of the $i$th oscillator in population 1 or 2, respectively. The populations are coupled internally with coupling strength $K_p$, whereas the interpopulation coupling is determined by $K$. The asymmetry is introduced into the model through the phase shift $0 \leq \alpha<\pi/2$ \cite{Kur84,SK86}. This permits the oscillators to synchronize to a frequency that deviates from the simple average of their natural frequencies. This behavior is common to many living systems such as mammalian intestine and heart cells \cite{Win80}. Moreover, such asymmetry appears in the phase reduction of nonisochronous oscillators \cite{PRK01,MB03} and Josephson junction arrays \cite{WS95,WCS96}, and it is used for modeling time delays \cite{PRK01} and information concerning synaptic connections in a neural network \cite{HI98}. The natural frequencies $\omega_i^{(1,2)}$ are considered to be distributed according to a density $g^{(1,2)}(\omega)$ of width $\gamma$, symmetric about the mean $\bar\omega^{(1,2)}$ and unimodal.

The phase coherence within each population is described by the complex order parameters $R^{(1,2)} e^{i\psi^{(1,2)}}=\frac{1}{N} \sum_{j=1}^{N}e^{i\theta_j^{(1,2)}}$, which permit to write the system (\ref{model0}) in terms of the mean field quantities $R^{(1,2)}$ and $\psi^{(1,2)}$    
\begin{equation}
\begin{split}
\dot\theta_i^{(1,2)}  = &\omega_i^{(1,2)}
- K_p \, R^{(1,2)} \sin(\theta_i^{(1,2)}-\psi^{(1,2)}+\alpha) \\ & 
- K \,   R^{(2,1)} \sin(\theta_i^{(1,2)}-\psi^{(2,1)}+\alpha).
\end{split}
\label{model}
\end{equation}

If the populations are uncoupled, i.e. $K=0$, each of them reduces to the well known Kuramoto model \cite{Kur84}. For a given $K_p$ this model exhibits a phase transition at a critical value of the frequency dispersal $\gamma_c$. For $\gamma>\gamma_c$ the oscillators rotate with their natural frequencies and $R^{(1,2)} \sim O(\sqrt{1/N})$, but for $\gamma<\gamma_c$ mutual entrainment occurs among a small fraction of oscillators giving rise to a finite value of the order parameter $R^{(1,2)}$. Thus, a cluster of locked oscillators emerges through a Hopf bifurcation of frequency $\Omega^{(1,2)}$ that, in general ($\alpha \neq0$), depends on the overall shape of $g^{(1,2)}(\omega)$ \cite{SK86}. The drifting oscillators arrange in a stationary distribution that does not contribute to the order parameters \cite{Kur84}.  Finally, for identical oscillators $\gamma=0$, each population fully synchronizes in-phase, $R^{(1,2)}=1$, for arbitrary small $K_p$.


When $K>0$ the two locked clusters begin to interact. If this interaction is similar to the frequency adjustment between two coupled oscillators, one expects mutual locking between these two clusters to occur via a saddle-node bifurcation at some $K=K_c$ \cite{PRK01}. Especially, for $\gamma=0$, synchronization should arise at ($\Delta\omega \equiv\bar\omega^{(1)}-\bar\omega^{(2)}$)   
\begin{equation}
K_c=\Delta\omega / (2\cos\alpha). 
\label{kc}
\end{equation}

\begin{figure}[tb]
\epsfig{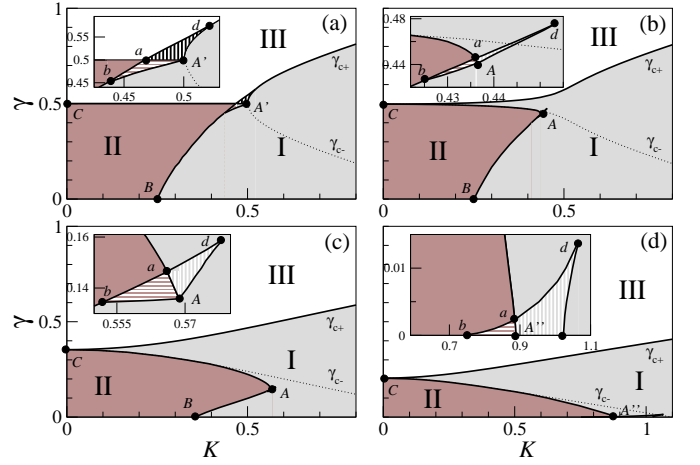}
\caption{$(K,\gamma)$ phase diagram of system (\ref{model0}) assuming Lorentzian frequency distributions, $\Delta \omega=0.5$, $K_p=1$, and  for different values of $\alpha$ (in rad.): (a) $\alpha=0$, (b) $\alpha=0.1$, (c) $\alpha=\pi/4$, (d) $\alpha=1.15$. Numerical stability boundaries ($N=1000$) are indicated as solid lines. Dotted lines represent analytical stability boundaries $\gamma_{c\pm}$ obtained from (\ref{eigenvalue}). Note that $\gamma_{c+}$ fully overlaps with numerical results. Region I: synchronization. Region II: coexistence. Region III: incoherence. Insets: bistability in the dashed regions around $A$ (see text).}
\label{diagrams}
\end{figure}

In the following we investigate the dynamics of (\ref{model}) in the full $(K,\gamma)$-parameter plane. In the thermodynamic limit a density function can be defined so that $\rho^{(1,2)}(\theta,t,\omega)d\omega d\theta$ describes the number of oscillators with natural frequencies in $[\omega,\omega+d\omega]$ and phase in $[\theta,\theta+d\theta]$ at time $t$. For fixed $\omega$ the distribution $\rho^{(1,2)}(\theta,t,\omega)$ of the phases $\theta$ is normalized to unity. The evolution of $\rho^{(1,2)}(\theta,t,\omega)$ obeys the continuity equation $\partial \rho^{(1,2)}/\partial t= - \partial(\rho ^{(1,2)} \dot\theta^{(1,2)})/\partial \theta$, for which  the incoherent state $\rho_0=(2\pi)^{-1}$ is always a trivial solution \cite{SM91}. The function $\rho^{(1,2)}(\theta,t,\omega)$ is real and $2\pi$-periodic in $\theta$ and therefore admits the Fourier expansion  $\rho^{(1,2)}(\theta,t,\omega)= \sum_{l=-\infty}^{\infty} \rho_l^{(1,2)}(t,\omega) e^{il\theta}$ ,
where $\rho^{(1,2)}_{-l}=\rho^{*(1,2)}_l$. Thus the order parameter can be written in terms of the Fourier components as $R^{(1,2)} e^{i\psi^{(1,2)}}= 2\pi \left< \rho^{*(1,2)}_{1} \right>$ (we use $\left< f^{(1,2)}(\omega) \right>$ to denote frequency average weighted with $g^{(1,2)}(\omega)$, respectively). Now, after inserting (\ref{model}) into the continuity equation, we obtain an infinite system of integro-differential equations for the Fourier modes
\begin{equation}
\begin{split}
\dot\rho^{(1,2)}_l=& -i \omega l \rho^{(1,2)}_l + \\ & l \rho^{(1,2)}_{l-1} \pi e^{i\alpha} \left(K_p \left< \rho^{(1,2)}_1 \right>+ K \left< \rho^{(2,1)}_1 \right> \right)  -\\ 
&  l \rho^{(1,2)}_{l+1}\pi e^{-i\alpha}  \left(K_p \left< \rho^{*(1,2)}_1 \right>+ K \left< \rho^{*(2,1)}_1 \right>\right).
\end{split}
\label{modes}
\end{equation}
The stability of $\rho_0$ can be analyzed by studying the evolution of a perturbed state $\rho^{(1,2)}(\theta,t,\omega)$ close to $\rho_0$ (note that the $\rho_l^{(1,2)}$ are then small quantities). Linearization of (\ref{modes}) reveals that the only potentially unstable modes are $l=\pm 1$ and hence $l=1$ has solution $\rho^{(1,2)}_1(t,\omega)=b^{(1,2)}(\omega)e^{\lambda t}+O(|\rho_l|^2)$. This leads to 
\begin{equation}
b^{(1,2)}(\omega)= \left(K_p\left<b^{(1,2)}(\omega)\right>+K \left<b^{(2,1)}(\omega)\right> \right) \frac{e^{i\alpha}/2}{\lambda+i \omega}.
\label{system}
\end{equation}
Considering the distribution of frequencies to be of Lorentzian type, $g^{(1,2)}(\omega)=(\gamma/\pi)[\gamma^2+(\omega-\bar\omega^{(1,2)})^2]^{-1}$, the self-consistent problem (\ref{system}) can be solved analytically. The stability of the incoherent state $\rho_0$ is then described by two pairs of complex conjugate eigenvalues, namely     
\begin{equation}
\lambda_\pm=-\gamma+\frac{K_p e^{i\alpha}}{2}\pm\frac{1}{2}\sqrt{K^2 e^{i 2 \alpha}-\Delta \omega^2} -i \bar\omega, 
\label{eigenvalue}
\end{equation}
with $\bar\omega\equiv(\bar\omega^{(1)}+\bar\omega^{(2)})/2$ for mode $l=1$, and the complex conjugate for $l=-1$. Imposing Re$(\lambda_\pm)=0$ defines explicitly the two critical curves $\gamma_{c\pm}(K)$ (see Fig.\ref{diagrams}). Each curve represents a Hopf bifurcation with frequency given by $\Omega_\pm \equiv -\mbox{Im}(\lambda_\pm)$. The curve max$(\gamma_{c+},\gamma_{c-})= \gamma_{c+}$ separates the region where the incoherent solution III is stable from the unstable regions I and II. 

The eigenmodes $\left<\rho^{(1,2)}(\theta,t,\omega)\right>$ near criticality are
\begin{equation}
\begin{split}
\binom{\left<\rho^{(1)}\right>}{\left<\rho^{(2)}\right>} = & \binom{1/2\pi}{1/2\pi} + Z_+(t) \binom{-i  z_0}{1} e^{i(\theta-\Omega_+ t)}+ \mbox{c.c.}+\\
& Z_-(t) \binom{1}{i z_0} e^{i(\theta-\Omega_- t)} +\mbox{c.c.} +O(|Z|^2),
\end{split}
\label{sol}
\end{equation}
where $Z_\pm(t)\equiv e^{\mbox{Re}(\lambda_\pm)t}$, and c.c. denotes the complex conjugate of the preceding term. The modulus of the number $z_0 \equiv  (\Delta\omega- \sqrt{\Delta\omega^2- e^{i2\alpha}K^2})e^{-i \alpha}/K$ is a weight for the fraction of frequencies $\Omega_+$ and $\Omega_-$ in populations 1 and 2, respectively.

\emph{The symmetric case, $\alpha=0$ [Fig.\ref{diagrams}(a)]}$.-$ Our eigenmodes (\ref{sol}) coincide with those of \cite{OK91} (replacing $\gamma$ by the noise intensity). From (\ref{eigenvalue}) the state III can become unstable in two different ways. When $K>\Delta\omega$ the transition III-I takes place through a single Hopf bifurcation and both populations synchronize to the same frequency $\Omega = \bar\omega$. The presence of a single macroscopic oscillation is denoted as region I. When $K<\Delta\omega$ the instability is through a \emph{degenerated} Hopf bifurcation. Both $(\lambda_\pm,\lambda_\pm^*)$ cross simultaneously the imaginary axis at $\gamma_{c\pm}=\gamma_c=K_p/2$ (line $\overline{CA'}$) and two macroscopic oscillations with frequencies $\Omega_\pm = \mp 1/2\sqrt{\Delta\omega^2-K^2} +\bar\omega$ emerge (note that a saddle-node bifurcation should take place at $A'$, i.e. $K_c=\Delta\omega$). The region of coexistence of two different macroscopic fields is labeled as II. 
The inset of Fig.\ref{diagrams}(a) shows how the saddle-node line $\overline{bad}$ crosses the two Hopf lines $\gamma_c$ at $a$, and joins the Hopf curve $\gamma_{c+}$ at $d$. Thus the Hopf bifurcation is supercritical all along $\gamma_{c+}$, except in $\overline{aA'd}$, where it is subcritical and hence a region of bistability between states III/I and II/I is observed.

\begin{figure}[tb]
\psfrag{om}{\Huge $\tilde \omega_i$} 
\psfrag{i}{\Huge $i$} 
\epsfig{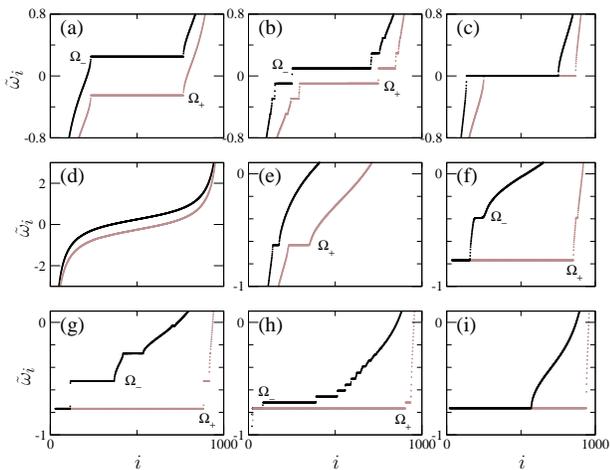}
\caption{Coupling-modified frequencies $\tilde\omega_i$ of populations 1 (black) and 2 (grey) as a function of the oscillator's index $i$: oscillator $i$ has natural frequency $\omega^{(1,2)}_i=\bar\omega^{(1,2)}+\gamma \tan{[(\pi/2)(2 i-N-1)/(N+1)]}$ (Lorentzian) ($\Delta\omega=0.5$, $\bar\omega=0$, $K_p=1$ and $N=1000$). First row: $\alpha=0$, $\gamma=0.4$ and (a) $K=0$, (b) $K=0.4$, (c) $K=0.41$. Second and third rows: $\alpha=\pi/4$, $K=0.53$ and (d) $\gamma=0.5$, (e) $\gamma=0.47$, (f) $\gamma=0.187$, (g) $\gamma=0.15$, (h) $\gamma=0.12$, (i) $\gamma=0.118$.}
\label{stairs}
\end{figure}

The coupling-modified frequencies of the individual oscillators $\tilde\omega_i^{(1,2)} = \lim_{t\rightarrow \infty} \theta_i^{(1,2)}/t$, provide a useful measure of synchronization: when $K=0$ ($\gamma<\gamma_c$) the frequency-locked oscillators in each population form a single plateau that is the only contribution to the order parameters (note that $|z_0|=0$ in (\ref{sol}), and hence $\psi^{(1)}=\Omega_- t$ and $\psi^{(2)}=\Omega_+ t$) [Fig.\ref{stairs}(a)]. By increasing $K$, some of the oscillators in populations 1 and 2 begin to lock in a second plateau at $\Omega_+$ and $\Omega_-$, respectively, according to (\ref{sol}) [Fig.\ref{stairs}(b)]. Hence, $R^{(1,2)}$ begin to oscillate with beating frequency $\Delta\Omega \equiv \Omega_ - -\Omega_+$. Interestingly new clusters synchronized to higher frequencies appear 
among those drifting oscillators with $\tilde\omega_i^{(1,2)}$ close to
\begin{equation}
\Omega_n=\Omega_- + n\Delta\Omega, \quad\mbox{where}\quad n=1,\pm 2,\pm 3....
\label{st}
\end{equation}
These plateaus $\Omega_n$, which are similar to Shapiro steps \cite{Sha63,Sak88}, are not explained by (\ref{sol}), but they can be understood from the fact that the drifting oscillators are simultaneously forced by the two order parameters (\ref{sol}). The plateaus grow in size and in number as $\Delta \Omega \rightarrow 0$  and hence they make a nonzero contribution to the order parameters that becomes important as the system approaches the saddle-node bifurcating line $\overline{Ba}$ from the region II. At this line the synchronized state I is reached, $\Delta \Omega=0$, and the steps abruptly disappear [Fig.\ref{stairs}(c)].

\begin{figure}[tb]
\epsfig{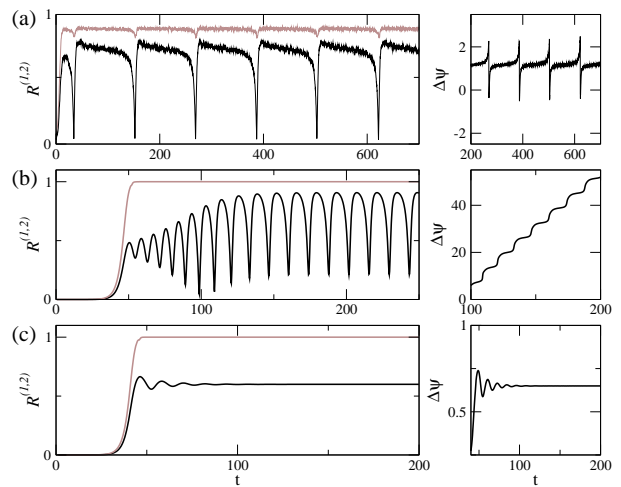}
\caption{Order parameters $R^{(1)}$ (black) and $R^{(2)}$ (grey) and phase difference $\Delta\psi$ as a function of time ($N=1000$, $\Delta \omega=0.5$, $\bar\omega=0$, $K_p=1$). At $t=0$ the phases were equally spaced in $(0,2\pi]$. (a) $\alpha=\pi/4$, $\gamma=0.12$, $K=0.53$ (region II) [Fig.\ref{stairs}(h)]; (b) and (c) represent regions II' ($\alpha=1.2$, $K=0.8$) and I' ($\alpha=1.2$, $K=1.05$), respectively ($\gamma=0$, see Fig.\ref{g0diagram}).}
\label{ops}
\end{figure}

\emph{The asymmetric case, $\alpha > 0$ [Figs.\ref{diagrams}(b,c,d)]$-.$} As $\alpha$ is increased from zero, the bifurcating lines $\gamma_{c+}$ and $\gamma_{c-}$ split due to the breaking of symmetry. Interestingly, the eigenmodes (\ref{sol}) do not reflect the asymmetry through the amplitudes $|z_0|$, but only through the different exponential growths $Z_\pm(t)$. Figs.\ref{stairs}(d-i) show the $\tilde\omega_i$ for $\alpha=\pi/4$ [Fig.\ref{diagrams}(c)], keeping $K$ constant and decreasing $\gamma$ continuously from region III.  We find that incoherence [Fig.\ref{stairs}(d)] always goes unstable through a single Hopf bifurcation $\gamma_{c+}$ (at
$\Omega_+$ in Fig.\ref{stairs}(e)) and nucleation first takes place mainly within population 2. The second Hopf bifurcation $\overline{CA}$ (at $\Omega_-$ in Fig.\ref{stairs}(f)) follows $\gamma_{c-}$ when the system is close enough to the incoherent state III. As $\gamma$ is decreased further, the system approaches the saddle-node bifurcation, $\overline{Bd}$, and an increasing number of oscillators in population 1 becomes entrained to the frequencies (\ref{st}) [Figs.\ref{stairs}(g,h)]. In consequence the order parameter $R^{(1)}$ oscillates  with a very large amplitude at frequency $\Delta\Omega$ whereas $R^{(2)}$ remains almost constant [Fig.\ref{ops}(a)]. The phase difference between the order parameters $\Delta\psi \equiv \psi^{(1)}-\psi^{(2)}$ reveals the presence of such clusters: $\Delta\psi$ [Fig.\ref{ops}(a)] is bounded despite the fact that the populations are not locked in frequency [Fig.\ref{stairs}(h)].

The bistability regions [Figs.\ref{diagrams}(b,c) inset]  are located around the intersection $a$ of the Hopf line $\overline{CA}$ with the saddle-node line $\overline{Bd}$. Within the region enclosed by $Aba$ the states I and II coexist, as in the $\alpha=0$ case. In contrast, the region enclosed by $Aad$ is surrounded only by the  state I and a new bistability between a small/large amplitude of the synchronized oscillation is observed.

With increase in $\alpha$,  the synchronization regions I and II become gradually smaller because as $\alpha\rightarrow \pi/2$ synchronization is increasingly inhibited due to frustration \cite{PRK01}.  At the same time, $|z_0|$ decreases indicating a lower degree of synchronization between the populations. This is in qualitative agreement with the approach of the saddle-node line $\overline{Bd}$ to the $\gamma=0$ axis [Figs. \ref{diagrams}(c,d)]. At the critical value $\alpha=\alpha^*$ the line $\overline{Bb}$ collides with the $\gamma=0$ axis and disappears (see Inset \ref{diagrams}(d)). Therefore, for $\alpha>\alpha^*$ synchronization between the macroscopic oscillations occurs generally when the oscillation of frequency $\Omega_-$ dies in the Hopf bifurcation $\overline{CA''}$.

\begin{figure}[tb]
\epsfig{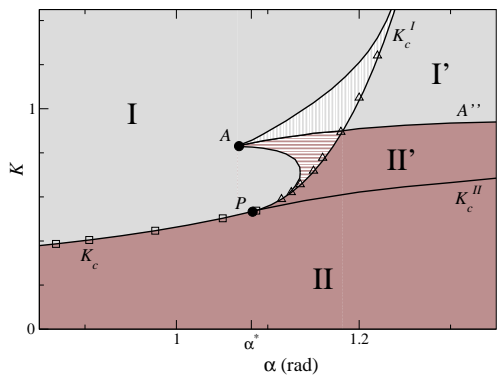}
\caption{Phase diagram $(K,\alpha)$ for $\gamma=0$, $\Delta\omega=0.5$. Boundaries $K_c$ and $K_c^{I}$ are obtained analytically from Eqs.(\ref{kc}) and (\ref{condition}), respectively, whereas the symbols $\square$ and $\vartriangle$ correspond to numerical results. All other boundaries are determined numerically. Regions I (synchronization) and II (drift) are characterized by $R^{(1,2)}=1$. Within regions I' ($\Delta\psi$ bounded) and II' ($\Delta\psi$ not bounded) $R^{(1)} <1$ whereas $R^{(2)}=1$ (see text). Dashed regions present bistability between states I and II' (horizontal dashes) and between I and I' (vertical dashes).}
\label{g0diagram}
\end{figure}

\emph{The limit $\gamma=0$ [Fig.\ref{g0diagram}]$.-$}  The transition point $B$ follows Eq.(\ref{kc}) as far as $\alpha <\alpha^*$ [Fig.\ref{diagrams}]. Since the oscillators within each population are identical, they synchronize in-phase, $R^{(1,2)}=1$, and the population's dynamics reduce to that of a system of two nonidentical oscillators. However, for $\alpha \geq \alpha^*$ the synchronization transition occurs via a Hopf bifurcation (line $A''$), and the behavior in each population is of higher complexity. As soon as $\alpha$ reaches the critical value $\alpha^*$ (point $P$), the curve $K_c$ splits into two bifurcating lines, $K_c^I$ and $K_c^{II}$, that enclose the new regions II' and I' where the order parameters are not synchronized [Fig.\ref{ops}(b)] and synchronized [Fig.\ref{ops}(c)], respectively. Within those regions the oscillators in population 1 are not in-phase synchronized, whereas the population 2 shows perfect in-phase entrainment [Figs.\ref{ops}(b,c)]. 

Finally we outline a linear stability analysis of the synchronized state I ($\gamma =0$) which confirms the loss of stability of the in-phase state of population 1. From (\ref{model0}), the phase difference between the populations is $\Delta\psi=\arcsin[{\Delta\omega/(2 K \cos{\alpha})}]$. Then linearization of (\ref{model0}) results in a matrix with $N-1$ eigenvalues $\mu_+$ and $N-1$ eigenvalues $\mu_-$ characterizing the stability of the in-phase synchronized state of populations 1 and 2, respectively 
\begin{equation}
\mu_\pm = -K_p \cos{\alpha}-K \cos{(\pm\Delta\psi+\alpha)}<0, 
\label{condition}
\end{equation}
and two eigenvalues $\mu_0=0, ~\mu_c=-2K \cos{\alpha}\cos\Delta\psi$  (note that $\mu_c=0$ leads to (\ref{kc})).  Since $\pi/2>\Delta\psi>0$, the condition (\ref{condition}) is only violated for the population 1. Thus
$\mu_+=0$ determines the boundary $K_c^{I}$ (and hence the point $P$) in very good agreement to numerics [Fig.\ref{g0diagram}].

Notice that with $K=0$ we recover the in-phase stability condition for a single population, $K_p \cos{\alpha}>0$. For $|\alpha|>\pi/2$ this state becomes unstable and reaches a neutrally-stable incoherent state. This issue has been the subject of a great deal of research in connection to the dynamics of devices consisting of Josephson junctions \cite{WS95}. In the present case, however, even for $|\alpha|<\pi/2$ the in-phase state in one population can be destabilized (population 1) or overstabilized (population 2) due to the interaction with the other population. The global stability properties of the states I' and II' in population 1 are interesting directions of further study: We stress that $R^{(1)}$ in Figs.~\ref{ops}(b,c) strongly depends on initial conditions and on perturbations, in contrast to $\Delta\psi$ and $R^{(2)}$. 

The mean field model (\ref{model0}) shows rich behavior despite of its simplicity, especially for $\alpha\neq0$. Beyond its importance for the theory of synchronization, oscillatory networks consisting of interacting subpopulations are common in neuroscience, and in general in many natural systems \cite{Win80}. For example, synchronization seems to be a central mechanism for neuronal information processing and for communication between different brain areas \cite{GKE+89,HI98}. This plays a crucial role in the pattern recognition and motor control tasks \cite{GKE+89}. In addition, the recordings of neuronal activity are usually taken in different brain regions, which constitute a network of interacting subpopulations of neurons \cite{GKE+89}. Thus, our study represents an important step into understanding macroscopic synchronization in complex network architectures.


We thank A.~J. G\'amez, G. Osipov, M.~G. Rosenblum, J. Schmidt, M.A. Zaks and D. Zanette for useful discussions.  This work was supported by  EU RTN 158 (E.M. and J.K.) and German VW-Stiftung (E.M. and B.B.).

\end{document}